\documentclass[reqno]{amsart}

\usepackage{amsmath}
\usepackage{amssymb}
\usepackage{amsthm}
\usepackage{mathtools}
\usepackage{physics}

\newcommand{\bbP}{\mathbb P}
\newcommand{\bbC}{\mathbb C}
\newcommand{\bbR}{\mathbb R}
\newcommand{\textdef}[1]{\textbf{\textit{#1}}}
\newcommand{\re}{\textnormal{Re}}
\newcommand{\im}{\textnormal{Im}}
\renewcommand{\phi}{\varphi}
\newcommand{\defeq}{\vcentcolon=}

\title{A Simple Constructive Proof of Wigner's Theorem}
\author{Daniel D.\ Spiegel}
\date{}
\address{University of Colorado Boulder\\ 
2000 Colorado Ave, Boulder, CO 80305}
\email{daniel.spiegel@colorado.edu}

\begin{document}

\begin{abstract}
This expository note presents a constructive proof of Wigner's theorem using only a few basic facts about Hilbert spaces, such as the existence of orthonormal bases and the Fourier decomposition of a vector. Our proof is based on a proof by Steven Weinberg found in the first volume of his series of textbooks on quantum field theory, but differs in a few places for the sake of greater simplicity and rigor.
\end{abstract}

\maketitle

\section{Introduction}
Wigner's theorem is a fundamental result in quantum mechanics that allows one to represent symmetry transformations of physical system by unitary or antiunitary operators on a Hilbert space, where the symmetry transformation is given by an isometry of the projective Hilbert space. Wigner's original proof \cite{wigner} is incomplete, but many complete proofs have emerged since then, see for example \cite{uhlhorn,bargmann,geher}. The proof presented in this paper is heavily based on a proof by Steven Weinberg \cite{weinberg}, but simplifies that proof and generalizes it by not requiring the symmetry to be surjective or the Hilbert space to be separable. Weinberg's proof is also not fully mathematically rigorous, and includes a possible division by zero which our proof circumvents. 

Let us present the necessary definitions and facts that will be used in the proof. First, we note that our Hilbert spaces will always be complex and our inner products will be linear in the second argument, in accordance with the physics convention. Given a complex Hilbert space $H$, we define its \textdef{projective Hilbert space} 
\[
\bbP H = \qty{\bbC \psi : \psi \in H \setminus \qty{0}},
\]
where $\bbC \psi = \qty{c \psi : c \in \bbC}$. Elements of $\bbP H$ are called \textdef{rays} and represent physical states. We define the \textdef{ray product}  $\ev{\cdot \,, \cdot}:\bbP H \times \bbP H \rightarrow \bbR$ as
\[
\ev{\bbC \psi_1, \bbC \psi_2} = \frac{\abs{\ev{\psi_1, \psi_2}}}{\norm{\psi_1}\norm{\psi_2}},
\]
where the angle brackets on the right denote the inner product on $H$. Physically, this represents a transition amplitude between two states. It is clear that this definition is independent of the representatives $\psi_1$ and $\psi_2$ of the rays. If $H$ and $H'$ are two complex Hilbert spaces, we define an \textdef{isometry} of projective Hilbert spaces as a map $\bbP H \rightarrow \bbP H'$, denoted $R \mapsto R'$, which preserves the ray product, i.e.\ $\ev{R_1, R_2} = \ev{R_1', R_2'}$ for any rays $R_1, R_2 \in \bbP H$. In particular, if we take normalized vectors $\psi_i \in R_i$ and $\psi_i' \in R_i'$ for $i = 1,2$, then
\begin{equation}\label{isometry}
\abs{\ev{\psi_1, \psi_2}} = \abs{\ev{\psi_1', \psi_2'}}.
\end{equation}
Symmetry transformations are implemented in quantum mechanics by bijective isometries, although we will not assume bijectivity of our isometry when we go to prove Wigner's theorem.

Given an isometry of projective Hilbert spaces, Wigner's theorem provides a corresponding linear or antilinear isometry of Hilbert spaces $U:H \rightarrow H'$. We define these terms as follows. A function $U:H \rightarrow H'$ between Hilbert spaces is an \textdef{isometry} (of Hilbert spaces) if 
\[
\norm{U\psi_1 - U\psi_2} = \norm{\psi_1 - \psi_2}
\]
for all $\psi_1,\psi_2 \in H$ and it is \textdef{antilinear} if 
\[
U(c_1\psi_1 + c_2 \psi_2) = c_1^*U\psi_1 + c_2^* U \psi_2
\]
for all $c_1, c_2 \in \bbC$ and $\psi_1, \psi_2 \in H$. We note that if $U:H \rightarrow H'$ is linear or antilinear, then it is an isometry if and only if $\norm{U\psi} = \norm{\psi}$ for all $\psi \in H$. 

The polarization identity:
\[
\ev{\psi_1, \psi_2} = \frac{1}{4} \sum_{k=0}^3 i^{-k} \norm{\psi_1 + i^k \psi_2}^2
\]
implies that a linear isometry $U:H \rightarrow H'$ satisfies 
\[
\ev{U\psi_1, U\psi_2} = \ev{\psi_1, \psi_2} 
\]
for all $\psi_1, \psi_2 \in H$, while an antilinear isometry $U:H \rightarrow H'$ satisfies
\[
\ev{U\psi_1, U\psi_2} = \ev{\psi_2, \psi_1}
\]
for all $\psi_1, \psi_2 \in H$.

Finally, our proof will use Bessel's inequality
\begin{equation}\label{Bessel}
\sum_\alpha \abs{\ev{\psi_\alpha, \psi}}^2 \leq \norm{\psi}^2,
\end{equation}
where $\qty{\psi_\alpha} \subset H$ is an orthonormal set and the sums here and below are understood in the sense of convergence of the net of finite partial sums.  Equality holds in (\ref{Bessel}) if and only if
\begin{equation}\label{Fourier}
\psi = \sum_\alpha \ev{\psi_\alpha, \psi} \psi_\alpha.
\end{equation}
We will also use the fact that every Hilbert space has a orthonormal basis $\qty{\psi_\alpha}$ and that any vector $\psi \in H$ may be expanded in the Fourier series (\ref{Fourier}) in this basis. Proofs of these facts can be found in many analysis texts, see for example \cite{rudin}. We are now ready to state and prove Wigner's theorem.

\section{Statement and Proof}

\noindent \textbf{Wigner's Theorem.}
{\itshape Let $H$ and $H'$ be complex Hilbert spaces and let $\bbP H \rightarrow \bbP H'$, $R \mapsto R'$ be an isometry. Then there exists an operator $U:H \rightarrow H'$ which is either a linear isometry or an antilinear isometry which respects the isometry of projective Hilbert spaces in the sense that
\begin{equation}\label{respectiso}
\psi \in R \,\,  \Longrightarrow \,\, U\psi \in R'.
\end{equation}
}

\begin{proof}
Fix an orthonormal basis $\{\psi_\alpha\}$ of $H$, with vectors belonging to rays $R_\alpha$. Let $\qty{\psi_\alpha'}$ be an arbitrary set of normalized vectors with $\psi_\alpha' \in R_\alpha'$. By (\ref{isometry}), these vectors are orthonormal
\begin{equation}\label{orthonormal}
\abs{\ev{\psi_\alpha',\psi_\beta'}} = \abs{\ev{\psi_\alpha,\psi_\beta}} = \delta_{\alpha \beta},
\end{equation}
where $\delta_{\alpha \beta}$ is the Kronecker delta. 
Furthermore, if $R \in \bbP H$, and $\psi \in R$ and $\psi' \in R'$ are arbitrary normalized vectors, then by (\ref{isometry}) and the saturation condition on Bessel's inequality we have
\[
\norm{\psi'}^2 = \norm{\psi}^2 = \sum_\alpha \abs{\ev{\psi_\alpha, \psi}}^2 = \sum_\alpha \abs{\ev{\psi_\alpha', \psi'}}^2,
\]
which implies that
\begin{equation}\label{primedFourier}
\psi' = \sum_\alpha \ev{\psi_\alpha', \psi'} \psi_\alpha'.
\end{equation}

Now we begin to construct $U$. Trivially, we define $U0 \defeq 0$. Next, let us single out some index, call the index $1$ for convenience, and define $U\psi_1 \defeq \psi_1'$. If $\dim H = 1$, then we define $U(c \psi_1) = c\psi_1'$ or $U(c\psi_1) = c^* \psi_1'$ for all $c \in \bbC$ and conclude the proof. Otherwise, for every other index $\alpha \neq 1$, we define
\[
\xi_{1\alpha}\defeq \xi_{\alpha 1} \defeq \frac{1}{\sqrt{2}}\qty(\psi_1 + \psi_\alpha),
\]
and let $S_\alpha$ denote the ray containing $\xi_{1\alpha}$. For any index $\beta$ and normalized vectors $\psi_\beta' \in R_\beta'$ and $\xi_{1\alpha}' \in S_\alpha'$, we know that
\begin{align*}
\abs{\ev{ \psi_\beta', \xi_{1\alpha}'}} = \abs{\ev{\psi_\beta, \xi_{1\alpha}}} = \left\{ \begin{array}{cl} 1/\sqrt{2} &: \beta = 1, \alpha\\ 0 &: \beta \neq 1, \alpha \end{array}\right..
\end{align*}
There are unique choices for the phase of $\xi'_{1\alpha}$ and $\psi_\alpha'$ such that
\begin{align*}
\ev{U\psi_1, \xi_{1\alpha}'} = \ev{\psi_\alpha', \xi_{1\alpha}' } = \frac{1}{\sqrt{2}}.
\end{align*}
We define $U\psi_\alpha$ and $U\xi_{1\alpha}$ to be the unique normalized elements of $R_\alpha'$ and $S_\alpha'$ that satisfy the above condition. By (\ref{primedFourier}), we know
\[
U\xi_{1\alpha} = \frac{1}{\sqrt{2}}\qty(U\psi_1 + U\psi_\alpha)
\]
We can see the linearity of $U$ beginning to take form. Note that the vectors $\qty{U\psi_\alpha}$ are orthonormal by (\ref{orthonormal}). 

If $\dim H = 2$, then the following paragraph should be skipped. Otherwise, we proceed as usual.

We will continue to define $U$ on several more specialized vectors in order to make defining $U$ on an arbitrary vector as simple as possible. Consider the vectors
\[
\eta_{\alpha \beta} \defeq \frac{1}{\sqrt{3}}(\psi_1 + \psi_\alpha + \psi_\beta), \qqtext{$1, \alpha, \beta$ distinct.}
\]
For any normalized $\eta_{\alpha \beta}'$ in the transformed ray, we know
\begin{align*}
\abs{\ev{U\psi_\gamma, \eta_{\alpha \beta}'}} = \abs{\ev{\psi_\gamma, \eta_{\alpha \beta}}} = \left\{ \begin{array}{cl} 1/\sqrt{3} &: \gamma = 1, \alpha, \beta\\ 0 &: \gamma \neq 1, \alpha, \beta \end{array}\right..
\end{align*}
We define $U\eta_{\alpha \beta}$ as the unique $\eta_{\alpha \beta}'$ with phase chosen so that the coefficient of $U\psi_1$ is real and positive; then by (\ref{primedFourier}) we have
\begin{align*}
U\eta_{\alpha \beta} = \frac{1}{\sqrt{3}} (U\psi_1 + c_\alpha U\psi_\alpha + c_\beta U\psi_\beta),
\end{align*}
where $\abs{c_\alpha} = \abs{c_\beta} = 1$. For $\gamma \in \qty{\alpha, \beta}$, the equality $\abs{\ev{U\xi_{1\gamma}, U\eta_{\alpha \beta}}} = \abs{\ev{\xi_{1\gamma}, \eta_{\alpha \beta}}}$ then implies
\begin{align*}
\abs{1 + c_\gamma} = 2.
\end{align*}
This implies that $c_\gamma = 1$, as can easily be shown with a moment of algebraic or geometric consideration.  Thus,
\[
U\eta_{\alpha \beta} = \frac{1}{\sqrt{3}}(U\psi_1 + U\psi_\alpha + U\psi_\beta).
\]

Next, we consider the vector 
\[
\xi_{\alpha \beta} = \frac{1}{\sqrt{2}}(\psi_\alpha + \psi_\beta), \qqtext{$\alpha, \beta$ distinct.}
\]
The case where $\alpha = 1$ or $\beta = 1$ reduces to the case we've already defined. Following our previous methods, we define $U\xi_{\alpha \beta}$ to be the unique element of the transformed ray such that
\[
U\xi_{\alpha \beta} = \frac{1}{\sqrt{2}}(U\psi_\alpha + c U\psi_\beta),
\]
where $\abs{c} = 1$. Then the equality $\abs{\ev{U\eta_{\alpha \beta}, U\xi_{\alpha \beta}}} = \abs{\ev{\eta_{\alpha \beta}, \xi_{\alpha \beta}}}$ implies $\abs{1+c} = 2$, which again implies $c = 1$.

We now begin to consider vectors with complex coefficients. Consider the vectors
\[
\phi_{\alpha \beta} = \frac{1}{\sqrt{2}}(\psi_\alpha + i\psi_\beta), \qqtext{$\alpha, \beta$ distinct.}
\]
We define $U\phi_{\alpha \beta}$ to be the unique vector in the transformed ray such that
\[
U\phi_{\alpha \beta} = \frac{1}{\sqrt{2}}(U\psi_\alpha + c U\psi_\beta),
\]
where $\abs{c} = 1$. The equality $\abs{\ev{U\xi_{\alpha \beta}, U\phi_{\alpha \beta}}} = \abs{\ev{\xi_{\alpha \beta}, \phi_{\alpha \beta}}}$ yields
\[
\abs{1+c}= \abs{1+i},
\]
which can be easily shown to imply either
\begin{subequations}
\begin{align}
c &= i \label{linear}\\
\text{ or \,\,} c &= -i. \label{antilinear}
\end{align}
\end{subequations}

The crux of the proof is to show that the same option of (\ref{linear}) or (\ref{antilinear}) must be taken for all $\phi_{\alpha \beta}$. First, observe that $\ev{\phi_{\alpha \beta}, \phi_{\beta \alpha}} = 0$ but $\abs{\ev{U \phi_{\alpha \beta}, U \phi_{\beta \alpha}}} = 1$ if different options are taken for $\phi_{\alpha \beta}$ and $\phi_{\beta \alpha}$. Thus, the same option must be taken for $\phi_{\alpha \beta}$ and $\phi_{\beta \alpha}$. This is all we must show if $\dim H = 2$.

If $\dim H \geq 2$, we consider next $\phi_{\alpha \beta}$ and $\phi_{\gamma \beta}$ where $\alpha$, $\beta$, and $\gamma$ are all distinct. Suppose $\phi_{\alpha \beta}$ obeys (\ref{linear}) and $\phi_{\gamma \beta}$ obeys (\ref{antilinear}). Consider the vector
\[
\psi = \frac{1}{\sqrt{3}}(\psi_\alpha + \psi_\gamma + i \psi_\beta)
\]
There exists a unique vector $\psi'$ in the transformed ray such that
\[
\psi' = \frac{1}{\sqrt{3}}(U\psi_\alpha + c_\gamma U\psi_\gamma + c_\beta U\psi_\beta),
\]
where $\abs{c_\gamma} = \abs{c_\beta} = 1$. By taking inner products with $U\xi_{\alpha \gamma}$ and $U\xi_{\alpha \beta}$ and using the isometry property (\ref{isometry}), we can conclude that $c_\gamma = 1$ and $c_\beta = \pm i$. If $c_\beta = i$, then equality of $\abs{\ev{\psi', U\phi_{\gamma \beta}}} = 0$ and $\abs{\ev{\psi, \phi_{\gamma \beta}}} = 2/\sqrt{6}$ gives a contradiction. On the other hand, if $c_\beta = -i$, then equality of $\abs{\ev{\psi', U\phi_{\alpha \beta}}} = 0$ and $\abs{\ev{\psi, \phi_{\alpha \beta}}} = 2/\sqrt{6}$ gives a contradiction, so we get a contradiction either way. Therefore, the same choice between (\ref{linear}) and (\ref{antilinear}) must be made between $\phi_{\alpha \beta}$ and $\phi_{\gamma \beta}$.

Finally, consider $\phi_{\alpha \beta}$ and $\phi_{\gamma \delta}$ for arbitrary indices $\alpha$, $\beta$, $\gamma$, $\delta$. We know the same choice must be made between $\phi_{\alpha \beta}$ and $\phi_{\delta \beta}$, as well as between $\phi_{\delta \beta}$ and $\phi_{\beta \delta}$, and also between $\phi_{\beta \delta}$ and $\phi_{\gamma \delta}$. Following this chain, we see that the same choice must be made between $\phi_{\alpha \beta}$ and $\phi_{\gamma \delta}$, as desired.

The work we've done up until now makes defining $U$ on an arbitrary nonzero vector $\psi$ easy. We expand $\psi$ as
\[
\psi = \sum_\alpha c_\alpha \psi_\alpha.
\]
Let $\alpha$ be an index such that $c_\alpha \neq 0$. We define $U\psi$ to be the unique normalized vector in the transformed ray such that the coefficient of $U\psi_\alpha$ is $c_\alpha$ if (\ref{linear}) is obeyed and $c_\alpha^*$ if (\ref{antilinear}) is obeyed. In other words, we define
\begin{equation}\label{generallin}
U\psi \defeq c_\alpha U\psi_\alpha + \sum_{\beta \neq \alpha} c_\beta' U\psi_\beta
\end{equation}
if (\ref{linear}) is obeyed, or
\begin{equation}\label{generalantilin}
U\psi \defeq c_\alpha^* U\psi_\alpha + \sum_{\beta \neq \alpha} c_\beta'^* U\psi_\beta 
\end{equation}
if (\ref{antilinear}) is obeyed, where $\abs*{c_\beta'} = \abs{c_\beta}$ for all $\beta$. This is consistent with the definitions we have made up until now. Now for any nonzero $c_\beta$, using the square of the isometry property (\ref{isometry}) with $\xi_{\alpha \beta}$ and $\psi$ leads to 
\[
\re(c^*_\alpha(c_\beta - c_\beta')) = 0,
\]
while using the square of the isometry property with $\phi_{\alpha \beta}$ and $\psi$ yields
\[
\im(c^*_\alpha(c_\beta - c_\beta')) = 0.
\]
Thus, $c^*_\alpha(c_\beta - c_\beta') = 0$, which implies that $c_\beta = c_\beta'$ since $c^*_\alpha \neq 0$. 

Thus, for every $\psi = \sum_\alpha c_\alpha \psi_\alpha \in H$ we see that either
\begin{align*}
U\qty(\sum_\alpha c_\alpha \psi_\alpha) = \sum_\alpha c_\alpha U\psi_\alpha \,\,\qqtext{ or }\,\,
U\qty(\sum_\alpha c_\alpha \psi_\alpha) = \sum_\alpha c_\alpha^* U\psi_\alpha,
\end{align*}
with the same choice taken across all $\psi \in H$. This implies that $U$ is either linear or antilinear. In either case, orthonormality of $\qty{\psi_\alpha}$ and $\qty{U\psi_\alpha}$ implies
\[
\norm{U\qty(\sum_{\alpha}c_\alpha \psi_\alpha)}^2 = \sum_\alpha \norm{c_\alpha}^2 = \norm{\sum_\alpha c_\alpha\psi_\alpha}^2,
\]
so $U$ is an isometry. This concludes the proof.
\end{proof}

\section{Conclusions}

We have presented a simple constructive proof Wigner's theorem, making as few assumptions as possible. In particular, we have considered an isometry between projective Hilbert spaces that do not necessarily come from the same Hilbert space, we have not assumed the Hilbert spaces to be separable, and we have not assumed the isometry to be bijective. As the proof uses only a few basic facts about Hilbert spaces that should be intuitive for physicists and well-known for mathematicians, we hope that this proof will be useful and accessible to all.

\section{Acknowledgments}

Thanks to Markus Pflaum for his advice and encouragement regarding this proof. This work was supported by the Department of Energy under Grant No. DE-FG02-91-ER-40672 and by the Center for Theory of Quantum Matter.

\bibliographystyle{ieeetr}
\bibliography{Wigner}
\end{document}